\documentclass[seceq]{ptptex}

\usepackage{graphicx}
\usepackage{wrapft}
\usepackage{color}
\usepackage[raggedright]{subfigure}

\title{Theory of heavy quark energy loss}

\author{Pol \textsc{Gossiaux}, J\"org \textsc{Aichelin}, and 
Thierry \textsc{Gousset}}

\inst{SUBATECH, Universit\'e de Nantes, EMN, IN2P3/CNRS
\\ 4 rue Alfred Kastler, 44307 Nantes cedex 3, France}

\abst{We briefly review some of the models and theoretical schemes established
to describe heavy quark quenching in ultrarelativistic heavy ions collisions.
Some lessons are derived from RHIC and early LHC data, especially as for the
contraints they impose on those models.}

\begin{document}

\maketitle

\section{Introduction} 
Jet-quenching of light hadrons is one of the key features observed in 
ultra-relativistic heavy ions collisions (URHIC) performed at RHIC 
($\sqrt{s}=200~A{\rm GeV}$). It has been recently confirmed in 
$\sqrt{s}=2.76~A{\rm GeV}$ lead-lead collisions at the LHC 
(see \cite{alice:2011} and references therein). This observation is usually 
interpreted as the consequence of the energy loss which affects the leading 
parton during its passage through a hot deconfined medium and is therefore 
crucial for scrutinizing its properties. In recent years, 
several theoretical schemes, based on the eikonal approximation, have been 
developed in order to describe this energy 
loss.\cite{BDMPS:97,GLV:00,Wiedemann:00,AMY}. Numerical evaluations
of these approaches turned out to be able to reproduce the rather flat 
momentum dependence of the nuclear modification factor ($R_{AA}$) observed at 
RHIC for pions at large $p_T$.
The driving parameters needed to obtain quantitative 
agreement,\cite{PHENIX:2008} however, overshoot pQCD predictions, sometimes 
by a factor 10\footnote{This large discrepancy should be taken with a grain 
of salt, 
as the optimal parameter vastly depends on the way the underlying 
medium is described in the model (see \cite{Enterria:2009}, 
for a recent discussion, as well as the contribution of S. Bass to these
proceedings).}. 
This might be taken as an indication that the 
fundamental theory is not able yet to describe the experimental results 
without introducing ad hoc parameters. 

This is one of the motivations for addressing on the same footing the 
quenching of jets consisting of leading heavy quarks (HQ), as they might help 
to better constrain the models. In this respect, the guiding concept is 
the so-called mass-hierarchy $\Delta E(g)>\Delta E(q)>\Delta E(c)>
\Delta E(b)$. The first inequality stems from the respective $SU(3)$
Casimirs of the gluons and quarks. The second is generally attributed to the 
dead cone effect\cite{Dokshitzer:2001} in the context of radiative energy loss 
but is present whenever the higher mass of the parton implies a reduction
of the formation time and hence of the radiated field. It is also found in collisional 
energy loss.\cite{Braaten:1991} 
Before the advent of the HQ data at RHIC, it was even advocated 
\cite{Djordjevic:2003} that HQ jets might be unquenched, but early 
$R_{AA}$ and elliptic flow ($v_2$) data of non-photonic single-electrons 
(NPSE) revealed that also HQ were strongly quenched in those collisions, 
nearly as much as light ones.\cite{PhenixandStar} Since then, 
some schemes designed for light quarks have been extended to heavy 
quarks\cite{Armesto1,Armesto2,WHDG} with the main conclusion that a common quantitative 
agreement between $R_{AA}(\pi)$ and $R_{AA}({\rm NPSE})$ can only be achieved 
if those NPSE stem exclusively from $c$-quarks. This is, however, 
incompatible with recent STAR measurements in $p-p$\cite{STAR:bcontent} and 
FONLL calculations\cite{Cacciari:2005} which indicate a significant component 
of $b$-quarks for $p_T>p_{T,{\rm cross}}\approx 5~{\rm GeV}/c$. Faced 
with this puzzle of lack of coupling of HQ with the QGP, several dedicated
models have flourished in the literature. In this contribution, we will 
review them shortly\footnote{See f.i. \cite{vHR:2009} for an extended 
review}. Then we will provide a tentative explanation why several models with 
various physical inputs are able to cope with data. Finally, we address the 
benefits of LHC and conclude.

\section{Models at RHIC: fragility and robustness}
\subsection{The two faces of heavy quark energy loss}
The first approach of HQ propagation in a hot medium was based on the 
Fokker-Planck 
equation with drag and diffusion coefficients evaluated
from collisional energy loss only.\cite{FP} 
This type of approach has survived up to now\cite{FP:recent}, although the 
physical content of the basic interaction has been extended by several 
authors (presented by increasing order of ``strong coupling'' content): 
A. Peshier\cite{Peshier} underlined the role of genuine running $\alpha_s$ in 
collisional energy loss. Calculations
implementing such running feature\cite{running1,running3} indeed find a smaller 
discrepancy with the $R_{AA}({\rm NPSE})$, sometimes
at the price of an additional cranking factor of the order of $2-4$.
Van Hees and Rapp were the first who have investigated the effect of 
possible heavy-light quark bound states in the $s$-channel.\cite{vHR:2005} 
Later, they have proposed a $T$-matrix description of HQ 
diffusion\cite{vHR:2008} based on 
a bona-fide generalization of the static $Q-\bar{Q}$ potential evaluated 
through lattice calculations. 
In both cases they obtain a good 
agreement with $R_{AA}$ as well as with $v_2$ data. Akamatsu et al.
\cite{Akamatsu:2009} have implemented the Langevin evolution of HQ in a 
hydrodynamical medium resorting to drag and diffusion coefficient evaluated 
in the strong coupling limit through AdS/CFT correspondence. Strictly 
speaking, all these FP/Langevin treatments 
applies for $p_T\lesssim m$. In this regime, HQ indeed behave as heavy 
particles surrounded by light degrees of freedom and their thermalization
time $\propto m$ differ significantly from that of light quarks. 

For ultra-relativistic HQ ($p_T\gg m$), the radiative energy loss becomes the 
dominant mechanism. The results of most of the eikonal schemes mentioned above
is the probability of radiation which presents significant fluctuations. 
In principle, this feature invalidates the FP treatment of HQ propagation
at high energy. 
In this regime, the mass of the quark acts  mostly as a collinear regulator. 
For the most energetic case ($E\gg m$), in-medium formation time of the high
energy gluons ($\sqrt{\omega/\hat{q}}$) exceeds the path 
length $L$. Then this scale regulates the radiation spectrum. 
As a consequence the average energy loss $\Delta E \propto L^2$
and the mass merely appears (if at all) through a logarithmic 
factor.\cite{BDMPS:97bis,PeigneSmilga}
It should be noted that the conversion of heavy quarks into heavy mesons
differ as well in these two regimes: for $p_T\lesssim m$, coalescence 
dominates because the probability to pick up a light quark from the medium 
at small relative velocity is large. For $p_T\gg m$, this 
probability is close to 0 and fragmentation becomes the dominant 
mechanism, with, however non trivial consequences on the $R_{AA}$ due to the 
presence of possible in-medium bound states.\cite{Sharma:09}

\subsection{Model fragility}
Although the approaches presented above differ vastly 
w.r.t their physical assumptions they are all able to cope with the 
NPSE $R_{AA}$ data measured by RHIC experiments, at the price of a
rescaling of the coupling parameter. 
\begin{wrapfigure}{r}{6.6cm} 
\includegraphics[width=6 cm]{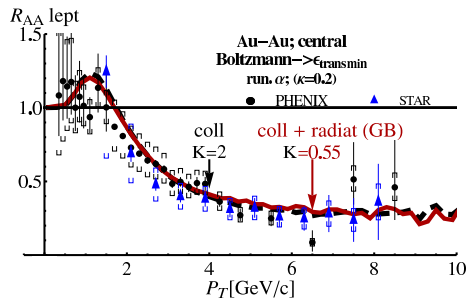}
\caption{$R_{AA}$ of non photonic single electrons for central Au-Au collisions
at RHIC energy; see \cite{Gossiaux:SQM09} for details.}
\label{rad_vs_col}
\end{wrapfigure} 
This questions our ability to 
achieve robust conclusions w.r.t. the basic mechanism at hand and 
hence the QGP properties.
 In\cite{Gossiaux:SQM09} we have implemented 
collisional as well as radiative energy loss in the same numerical framework 
(MC$\alpha_s$HQ) and confirmed this ``model fragility'' (see fig.
\ref{rad_vs_col}). A finer analysis reveals that the $R_{AA}$
observable is mostly sensitive to the energy loss spectrum $P(\omega)$ at low 
values of the energy loss $\omega$. As pointed out by Baier et 
al\cite{Baier:2001}, this is due to 
the rather stiff initial $p_T$-distribution which makes the sequence of many  
small losses more probable than a single process involving large energy 
loss. This explains why both spectra (see \cite{Vogel:2011} for an 
illustration) lead to similar quenching although the average 
$\Delta E_{\rm rad}>\Delta E_{\rm col}$.   

\subsection{``Robust'' contact with lattice calculations}
According to our analysis, all HQ observables at RHIC can be explained 
with a simple $\Delta E\propto L$ law, an observation in favor of local 
processes.
 \begin{wrapfigure}{r}{6.6cm} 
\includegraphics[width=6.cm]{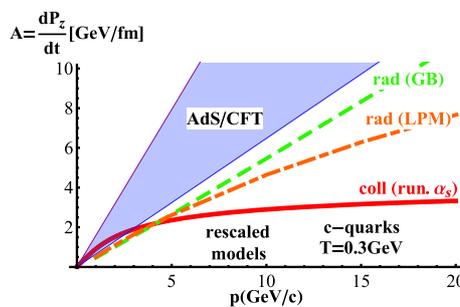}
\caption{Drag coefficient of various models incorporated in 
MC$\alpha_s$HQ (running $\alpha_s$ collisional, incoherent and 
coherent radiation), with coupling parameter rescaled in order to match 
the NPSE RHIC data.}
\label{fig_drag}
\end{wrapfigure} 
On fig. \ref{fig_drag}, we show the drag coefficient $A$ from 
the different microscopic energy-loss models implemented in our 
MC$\alpha_s$HQ framework. Although these models vary largely, their values 
for $A$ nicely approach a rather unique value for $p_T\lesssim 
5~{\rm GeV}/c$ once
the coupling parameter (here the interaction rate) is rescaled in order
to match RHIC NPSE data. From this we extract ``robust'' values of the 
relaxation coefficient $\eta=\lim_{p\rightarrow 0}\frac{A(p)}{p}$ of
$0.45\pm 0.15~c/{\rm fm}$ at $T=200~{\rm MeV}$ and
$0.75\pm 0.25~c/{\rm fm}$ at $T=300~{\rm MeV}$. This yields 
a spacial diffusion coefficient $D_s$ of 
$2\pi T D_s=1.9\pm 0.5$ at $T=200~{\rm MeV}$ and of
$2\pi T D_s=2.55\pm 0.65$ at $T=300~{\rm MeV}$, for both $c$ and $b$ quarks,
in quite good agreement with recent lattice calculations.\cite{Ding:2011} 
This first successful contact with lattice should encourage us to seek for 
alternative observables able to discriminate between various models for 
$p_T\gtrsim 10~{\rm GeV/c}$.

\subsection{The role of the elliptic flow}
In principle, the $v_2$ observable helps in constraining the models. At low 
and intermediate $p_T$, it reflects the collectivity acquired by heavy quarks and  
develops constantly with time up to the end of the transition.\cite{Rapp_2008} It is thus
 more sensitive to the QGP evolution than the $R_{AA}$ which 
saturates earlier. Nevertheless, it has been shown recently\cite{Gossiaux:2011} that two
energy loss models with drag factor differing by a factor as large as two were both 
able to reproduce the experimental $R_{AA}$ and $v_2$ once they are imbedded
in different QGP evolution models chosen by the respective authors.\cite{vHR:2005,running2} 
This clearly points towards the need of performing joint analysis of bulk QGP properties
and HQ observables to achieve significant progress, as initiated in\cite{TRenk} 
for the case of light hadrons. At large $p_T$, one expects some $v_2$ as well, due to
the path length difference along both principal directions of the QGP, understood as
a source of quenching.\cite{Voloshin:0000} Although this observable could be 
useful to assess the path length dependence of HQ quenching (see\cite{Armesto2} for 
prediction within the ASW model), present 
$p_T$ range available at RHIC seems to offer little discriminating power for this 
purpose.

\section{Benefits from early LHC}  
URHIC performed at LHC offer larger possibilities to discriminate between 
various models and theoretical schemes, due to wider $p_T$ range and due to
the possibility to measure  $D$ and $B$-mesons observables separately from 
the very first runs on.
Theoretical predictions\cite{LHCpred,Sharma:09} for the $R_{AA}$ of 
$D$-mesons range between 0.2 and 0.4 for $p_T\gtrsim 10~{\rm GeV}/c$. First 
ALICE results on $D$ mesons in Pb-Pb collisions\cite{Dainese:2011}
confirmed the HQ quenching found at RHIC, with $R_{AA}\approx 0.3\pm 0.15$ for 
$p_T\in[5~{\rm GeV}/c,10~{\rm GeV}/c]$. 
\begin{figure}
\centerline{
\includegraphics[width=6 cm]{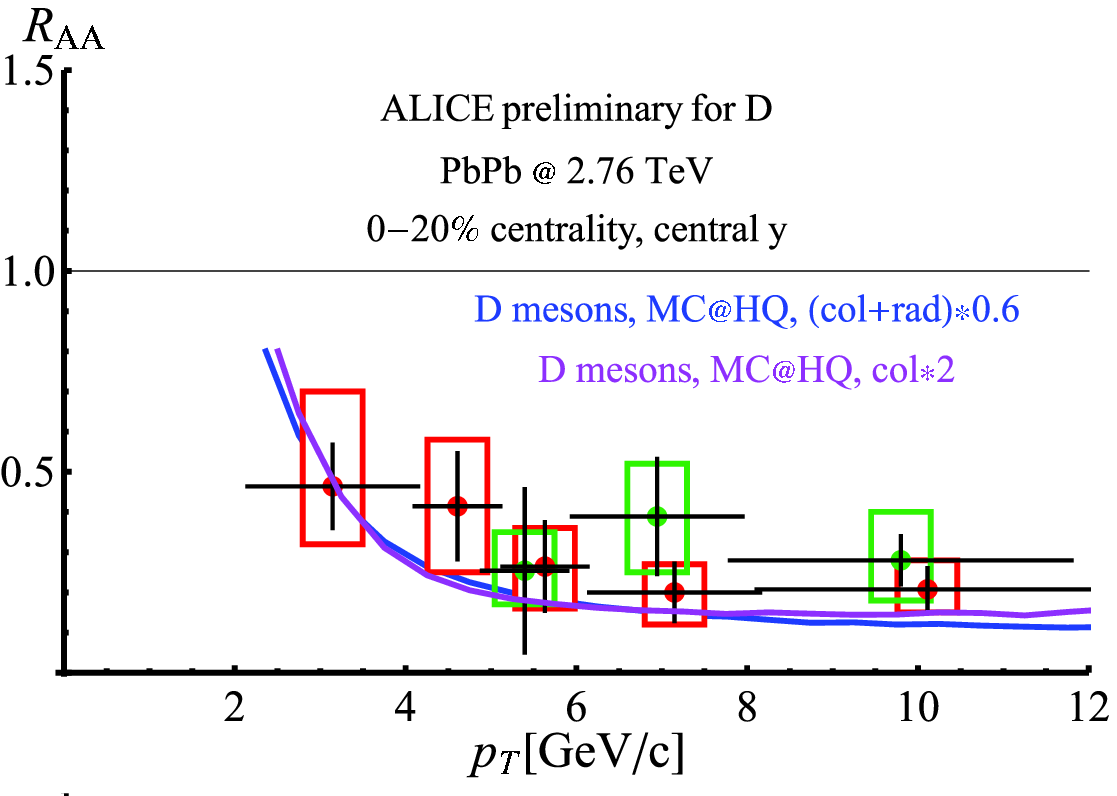}
\hspace{1cm}
\includegraphics[width=6 cm]{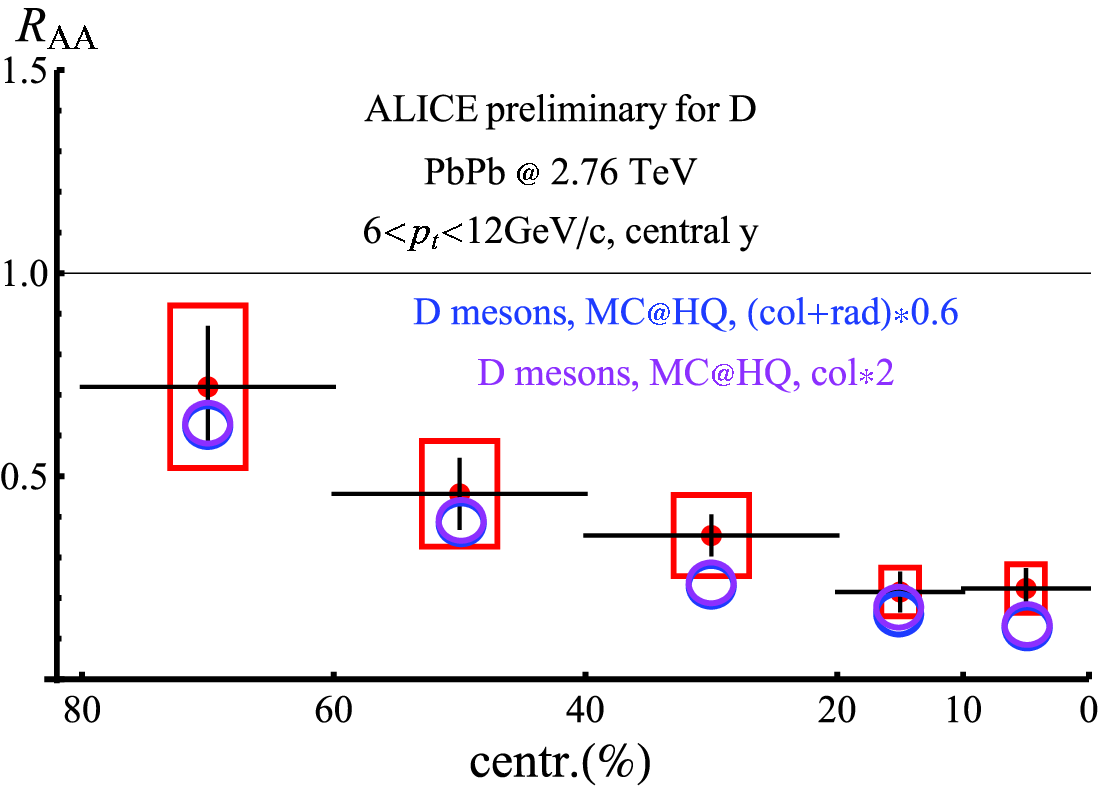}}
\caption{Quenching of D mesons in PbPb collisions at 2.76 TeV: preliminary 
ALICE data\cite{Dainese:2011} vs collisional + radiative energy loss 
implemented in MC$\alpha_s$HQ.}
\label{D_at_LHC}
\end{figure}

\begin{figure}[h]
\begin{minipage}[c]{210pt}
\includegraphics[width=7. cm,height=4.25cm]{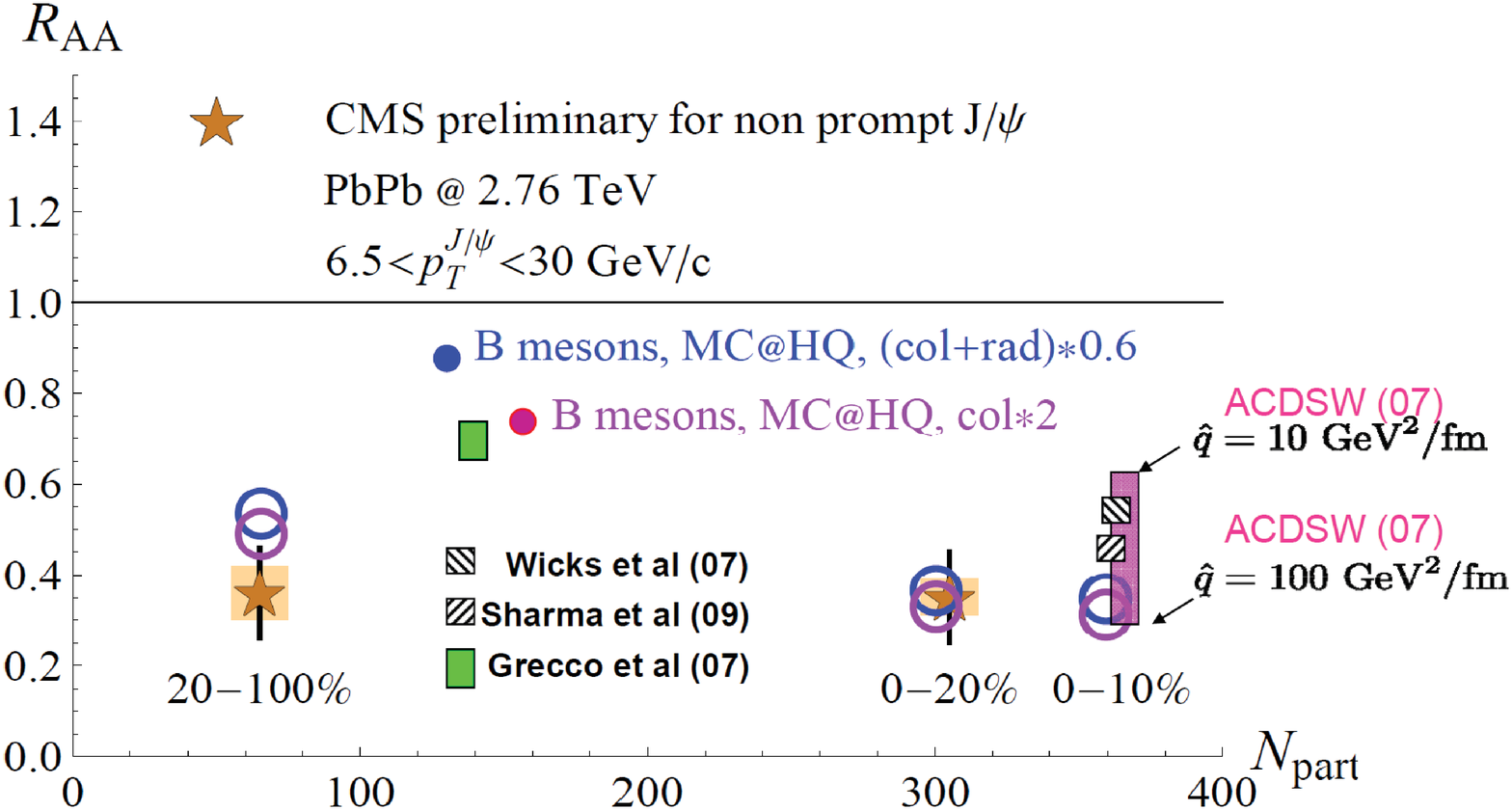}
\end{minipage}
\begin{minipage}[c]{210pt}
\includegraphics[width=6.5 cm]{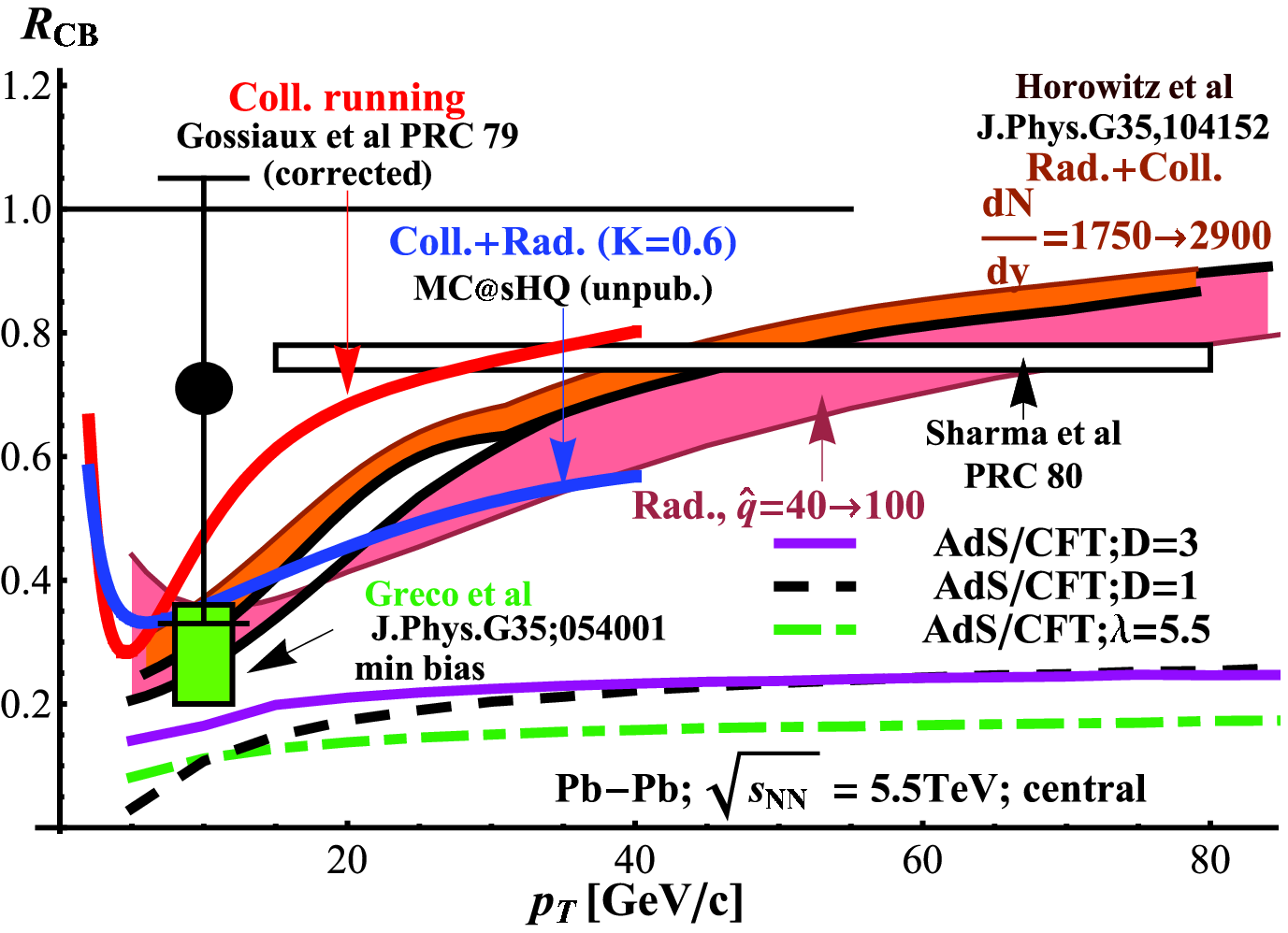}
\end{minipage}
\caption{Left: quenching of B mesons in PbPb collisions at 2.76 TeV 
(identified to the quenching of non-prompt $J/\psi$ measured by 
CMS\cite{Dahms:2011} (preliminary data) vs various model predictions. 
Right: $R_{CB}$ predictions
vs experimental ratio extracted from ALICE data on $D$-mesons and 
CMS data on non-prompt $J/\psi$ in central PbPb collisions (black disk).; caution: the 
prediction of Greco et al. was made for minimum bias collisions.}
\label{B_at_LHC}
\end{figure}

On fig. \ref{D_at_LHC}, we provide
the comparison between data and calculations performed with the same 
elementary reaction cross section as for the RHIC energy. We only 
adapted the plasma expansion (simulated with the hydrodynamical
code of Kolb-Heinz \cite{KolbHeinz}) to obtain $\frac{dN_{\rm ch}}{dy}=1600$ at the chemical 
freeze out
and the initial $p_T$-distribution (taken according to FONLL 1.3.2),
which is harder than at the RHIC energy\footnote{This explains why 
the medium appears as less opaque to HQ propagation although it is denser.} 
A satisfactory agreement is obtained although the $p_T$-range of the present 
data is not sufficient to disambiguate between collisional and radiative 
energy loss. 
Apart from the $p_T$ dependence of $R_{AA}$ and $v_2$, the 
various models exhibit a rather rich mass-dependence which can be probed
by addressing simultaneously the light vs heavy or $c$ vs $b$ observables.
The quenching of $B$ mesons has not been measured directly yet but can be
{\em bona fide} assimilated to the quenching of non-prompt $J/\psi$ measured
by CMS\cite{Dahms:2011}; this is legitimate due to the flat shape of $R_{AA}$ 
at high $p_T$. On fig. \ref{B_at_LHC}, we show the comparison between these
data and some collected predictions.\cite{LHCpred,Sharma:09} Most of the 
predictions
show a lack of quenching as compared to the data. If confirmed, this would 
indicate that history might repeat itself (large quenching of $c$ quark 
unpredicted and observed at RHIC; large quenching of $b$ quark unpredicted 
and observed at LHC) and would question our present ``understanding'' of
HQ propagation in hot media. 
On fig. \ref{B_at_LHC} (right), we present various predictions for the 
$R_{CB}$
ratio (defined as $R_{AA}(D)/R_{AA}(B)$) as a function of $p_T$, together
with the experimental point obtained by 
$R_{AA}(B)=R_{AA}({\rm non-prompt~}J/\psi)$ measured by CMS. Although
the rather large experimental errors prevent us presently from 
any firm conclusion, models incorporating rather large mass dependence like
AdS/CFT (drag coefficient $\propto m^{-1}$) seem to be disfavored by the 
data, although the analog of the detailed in-medium evolution of 
\cite{Akamatsu:2009} is lacking at LHC.

\section{Conclusions and prospects}
We have argued that the limited dynamical range of observables at RHIC hinders
the discrimination between the various models proposed for HQ 
evolution in hot media. 
We have shown that it was nevertheless possible 
to extract some basic properties of the QGP-HQ interaction 
-- like the drag coefficient at low momentum -- in a rather ``robust'' 
way, once those models are rescaled in order to match $R_{AA}$ and $v_2$ data.
These values are in good agreement with present lattice calculations. We have discussed
the recent LHC results pertaining to $D$-mesons and non-prompt 
$J/\psi$ from $B$ meson decay. In particular, we found that most of the
predictions seem to lack quenching for the $b$-quark. Whether it is just a 
question of fine-tuning of the present models or the sign of a more 
fundamental misunderstanding of the physics requires refined data and access 
to the full $p_T$ range of observables like $R_{CB}$. They will come in the 
next few years. Finding $R_{AA}(q)\approx
R_{AA}(c)\approx R_{AA}(b)$ at mid $p_T$ could be the sign that the 
gluon formation-time which is found to be an increasing function of the 
radiator's inverse mass\cite{Arnold:2008} could be bounded from above by 
another scale of the problem as for instance the gluon absorption length in 
the medium.\cite{Damping}

\section*{Acknowledgments}
This work was performed under the ANR research program ``hadrons @ LHC'' 
(grant ANR-08-BLAN-0093-02) and the PCRD7/I3-HP program TORIC.

%

\end{document}